\documentclass{article}
\usepackage{spconf,amsmath,graphicx}
\usepackage{makecell}

\usepackage{enumitem}
\setlist{nosep, leftmargin=14pt}

\usepackage{mwe} 

\title{A Radiogenomics Pipeline for Lung Nodules Segmentation and Prediction of EGFR Mutation Status from CT Scans}
\name{Ivo Gollini Navarrete, Mohammad Yaqub} 
\address{Mohamed bin Zayed University of Artificial Intelligence, Abu Dhabi, United Arab Emirates. }

\begin{document}
%
\maketitle
\begin{abstract}
Lung cancer is a leading cause of death worldwide. Early-stage detection of lung cancer is essential for a more favorable prognosis. Radiogenomics is an emerging discipline that combines medical imaging and genomics features for modeling patient outcomes non-invasively. This study presents a radiogenomics pipeline that has: 1) a novel mixed architecture (RA-Seg) to segment lung cancer through attention and recurrent blocks; and 2) deep feature classifiers to distinguish Epidermal Growth Factor Receptor (EGFR) mutation status. We evaluate the proposed algorithm on multiple public datasets to assess its generalizability and robustness. We demonstrate how the proposed segmentation and classification methods outperform existing baseline and SOTA approaches (73.54 Dice and 93 F1 scores).
\end{abstract}
\section{Introduction}
\label{sec:intro}
Lung cancer remains as the deadliest type of cancer, and in 2020 it represented 1.8 million deaths \cite{sung2021global}. Death rates are decreasing thanks to significant improvements in cancer treatment. However, long-standing prevention and early detection are critical to improving patient healthcare outcomes further. Doctors utilize a combination of invasive and non-invasive methods to define tumor characteristics and decide on a treatment strategy. Medical imaging such as computed tomography (CT) and positron emission tomography (PET) are used to obtain a detailed image of the chest area and determine metastasis status. On the other hand, molecular techniques detect mutated oncogenes and tumor suppressors for personalized treatment. Despite the promising potential of molecular techniques, biopsy presents several limitations: risk of patient complications, tumor heterogeneity, and the high cost of extraction/sequencing \cite{wong2020radiogenomics}.

Radiogenomics is an interdisciplinary research area that uses machine learning (ML) methodologies to link instructive image information to valuable genomic data. These techniques can potentially improve the sensitivity and specificity of diagnostic imaging \cite{wong2020radiogenomics}. For example, Brown \textit{et al.} \cite{brown2008use} detected glioma genetic signatures employing magnetic resonance imaging (MRI) scans, and Segal \textit{et al.} \cite{segal2007decoding} were able to infer molecular properties of liver lesions in CT scans. Motivated by the growing interest in radiogenomics applications, this work proposes a radiogenomics pipeline for non-small cell lung cancer (NSCLC), which corresponds to around 80\% and 85\% of lung cancers worldwide \cite{Siegel2022}.

We present a mixed-supervised 3D architecture for nodule segmentation. The proposed segmentation method introduces 1) an attention module to help guide the model localize the tumor accurately, 2) a recurrent spatio-temporal block to capture slice interdependencies, and 3) an augmentation step to overcome the problem of data scarcity. In addition, deep features are extracted, processed for dimensionality reduction, and passed into ML classifiers to predict the epidermal growth factor receptor (EGFR) mutation status. 
Ultimately, this pipeline can provide crucial visual cues and measurements from the segmentation, with predictive genomic information to help clinicians perform a reliable and non-invasive assessment of treatment before the biopsy.

\section{Related Work}
\label{sec:related_work}

Manual segmentation of the volume of interest (VOI) is the most precise method, but it needs experienced personnel, is time-consuming, and suffers from interobserver variability. On the other hand, automated processes are time efficient, but they easily overfit, need large dense-labeled datasets, and have difficulty covering various nodule appearances. Some approaches try to alleviate the use of large amounts of annotated data  by combining non-annotated data or weak labels like bounding boxes \cite{shah2018ms}. More recently, transfer learning using artificial datasets generated with a generative adversarial network (GAN) has been assessed \cite{nishio2021lung}.

Traditional radiomics approaches extract significant radiomic features that lead to efficient models for the prediction of gene mutation status. ML algorithms have been used to predict EGFR and Kirsten rat sarcoma virus (KRAS) biomarkers after tumor segmentation in CT and PET-CT images \cite{koyasu2020usefulness}. DL approaches have demonstrated higher performance over traditional methods \cite{moreno2021radiogenomics}, and state-of-the-art (SOTA) approaches use auxiliary tasks or multitasking to improve performance \cite{gui2022air}. 

\section{Materials and Methods}
\label{sec:materials_methods}

\subsection{Datasets}
The segmentation framework is trained and evaluated on three public datasets. The ``NSCLC-Radiomics" \cite{aerts2014decoding} (RAD) dataset contains 422 patients with available CT and segmentation masks, forty-two removed due to poor quality segmentation, while the ``Medical Decathlon Lung" \cite{simpson2019large} (MSD) dataset presents 63 patients. Following the radiogenomics pipeline, the ``NSCLC-Radiogenomics" \cite{bakr2018radiogenomic} (RADGEN) dataset is used as a segmentation test set and for the downstream classification task. It comprises 211 patients with available CT images. Tumor segmentations are available for 144 patients, from which EGFR mutation status is known for 117 (94 wildtypes and 23 mutant).

\begin{figure}[h]
    \centering
    \includegraphics[scale=0.25]{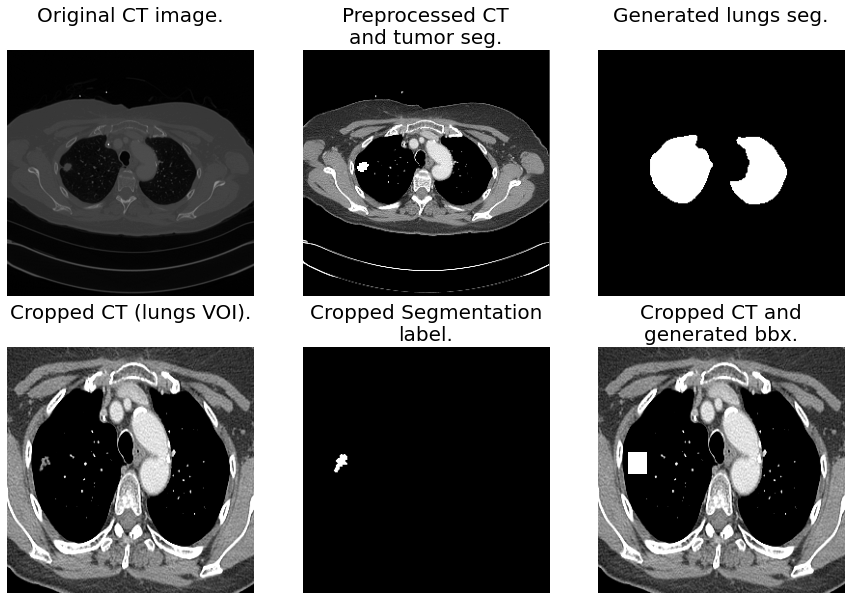}
    \caption{Illustrated example of data preprocessing, with corresponding lung VOI generated
mask and tumor bounding box.}
    \label{fig:preprocess}
\end{figure}

The three datasets undergo the same preprocessing: voxel intensity clipped to the range [-200, 250], resampling to an anisotropic resolution of $1\times1\times1.5$ $mm^3$, Lungs VOI crop utilizing U-Net(R231) pretrained weights \cite{hofmanninger2020automatic}, resize to 256×256×256 using b-spline for the image and nearest interpolation for the segmentation, and generation of 3D bounding boxes using the tumor segmentation mask. Figure \ref{fig:preprocess} illustrates the data preprocessing, including generated lung masks and tumor bounding boxes.

\subsection{Segmentation}
\begin{figure*}[ht]
    \centering
    \includegraphics[width=\textwidth]{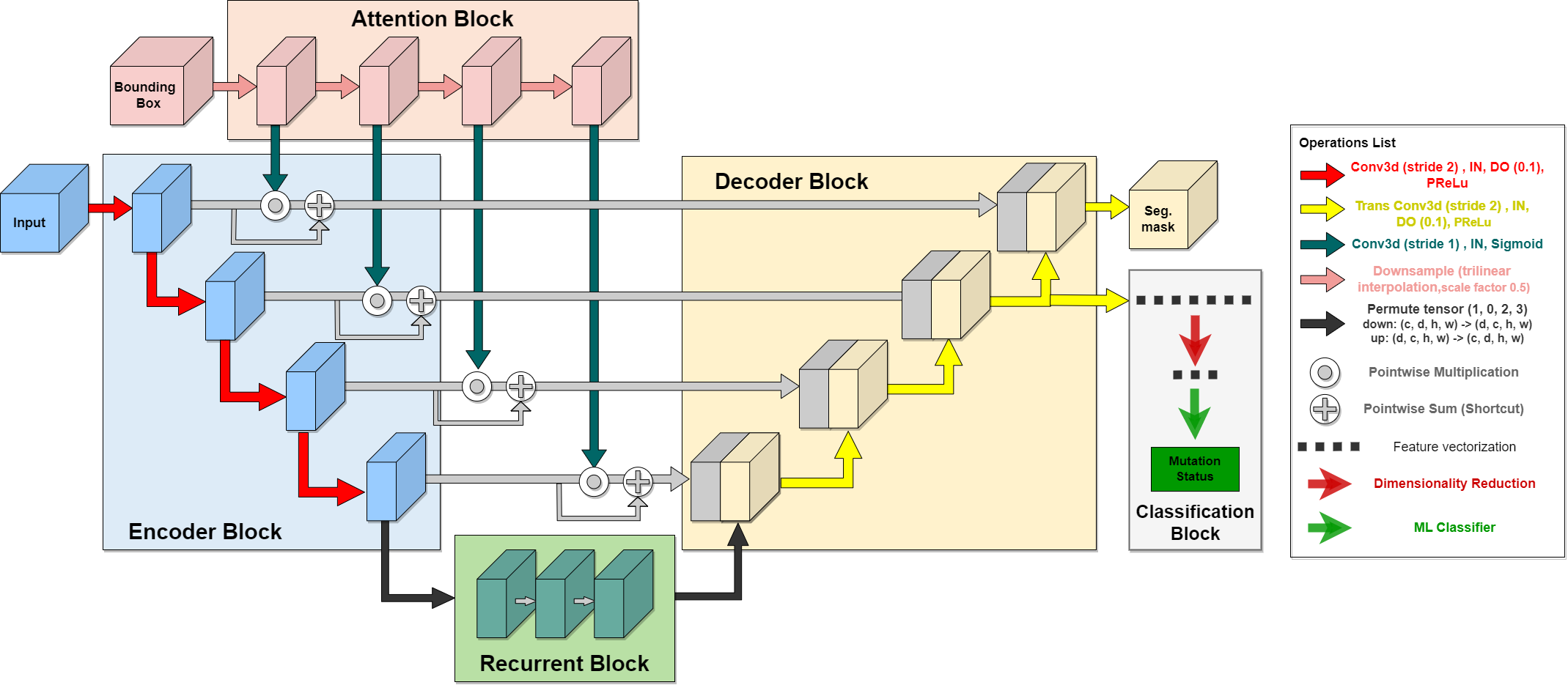}
    \caption{The proposed ``Recurrecen-Attention Segmentor" (RA-Seg) generates a tumor mask with a \textit{U-Net like} structure, the introduction of an \textit{Atenttion block} to recalibrate the skip connections, and a \textit{Recurrent Block} to capture slice interdependencies of the CT. Classification block branches from the \textit{Decoder block} to process deep features and output the biomarker prediction.}
    \label{fig:architecture}
\end{figure*}

The proposed architecture combines the core concepts from three models: volumetric U-Net \cite{cciccek20163d}, Recurrent 3D-DenseUNet \cite{kamal2020lung}, and an organ-to-lesion (O2L) module inspired from \cite{sun2020teacher}. Figure \ref{fig:architecture} illustrates the proposed model. MONAI's implementation of U-Net architecture is used as the backbone because of its flexibility in introducing ad-hoc designs.

Each step of the \textit{Encoder Block} is composed of a 3D-Convolutional block that extracts a series of features (64, 128, 256, 512) for context understanding. \textit{Decoder Block} concatenates the features coming from the corresponding encoder block and upsampled features from the previous decoder block to enable precise localization. Instance Normalization (IN) and spatial dropout with a rate of 0.1 are employed to alleviate overfitting. Parametric rectifying linear units (PReLU) are used to learn better activation as they are found to provide better performance than ReLU \cite{DelvingDeepintoRectifiers}.

The \textit{Recurrent Block} transforms the high-level features from the encoder block and captures the interdependencies in a more fine-grained spatio-temporal feature space \cite{kamal2020lung}. It comprises three Convolutional Long Short-Term Memory (ConvLSTM) layers with a kernel size $3\times3$.

A progressive weighting procedure enhances the shallow layers using an attention-to-tumor (a2t) module to help the model localize the tumor. At each step of the \textit{Attention Block}, features passed on the skip connection are recalibrated by

\begin{equation}
    f_{out} = f_{in} + \sigma(conv(b_{t})) \odot f_{in}
\end{equation}

where $conv(b_{t})$ denotes 3D-Convolution layers followed by $\sigma$ sigmoid activation function. A point-wise multiplication, represented by $\odot$, merges the input and bbx features to mask out features from the non-tumor region. A shortcut connection reintroduces non-tumor region features while enhancing the tumor region.

The following augmentations are used ``on the fly" to improve generalizability: Vertical Random Flip (50\% probability), Horizontal Random Flip (50\% probability), Random Rotation over the three axes (range -10 to 10 degrees), Random Shear over the three axes (range -10 to 10 degrees), and Random Scale over the depth axis (factor between 0.9 and 1.1). 

All experiments follow a 5-fold cross-validation scheme using an Adam optimizer with a learning rate of $10^{-3}$ until \textit{Dice Coefficient Score} (DSC) validation convergence. Optimization of parameters uses MONAI's \textit{Dice-Cross-Entropy Loss}, a weighted sum that takes advantage of the distribution-based Cross-Entropy (CE) loss and region-based Dice loss \cite{jadon2020survey}.




\subsection{Classification}
Segmentation is used as an auxiliary task to extract features. High-level features are extracted from the second to the last step of the \textit{Decoder Block} as this was found to produce the best classification results. Extracted features undergo a \textit{mean} operation to reduce the dimension size, and a \textit{flatten} function generates a vector with $8192$ deep features. A heavier model is avoided using ML methods to perform binary classification. 

Principal component analysis (PCA) and linear discriminant analysis (LDA) were tested to keep only meaningful properties. LDA dimensionality reduction uses the \textit{singular value decomposition} (svd) solver, while PCA determines components that explain \textit{95\% of the variance}. The ML methods used for binary classification are as follows.
(1) Quadratic discriminant analysis (QDA) uses a quadratic decision surface where a Gaussian distribution models the likelihood of each class \cite{srivastava2007bayesian}. 
(2) Decision Tree (DT) is a model that comprises discriminant functions to divide a dataset's feature space into pure, single-class subspaces \cite{biau2016random}. 
(3) Random Forest (RF) overcomes the overfitting problem of a DT by ensembling a series of randomized individual trees \cite{biau2016random}. 
(4) C-Support Vector Classification (SVC) maximizes the margin between the boundary hyperplane(s) and the closest data point on each side; dot products are replaced with a nonlinear kernel function to allow differentiation of non-linearly separable space set \cite{novakovic2011c}. 

Scikit-Learn's ``GridSearchCV" is implemented to find the best parameters for each model. Parameters search takes into consideration label imbalance by maximizing the F1-macro score. QDA hyperparameters are the default, DT utilizes the \textit{gini impurity} criterion, RF uses \textit{200 trees}, and SVC uses \textit{sigmoid kernel}.

\section{Results \& Discussion}
\label{sec:results_discussion}

\subsection{Segmentation}
As seen in Table \ref{tab:Final_arch}, the proposed method outperforms the baseline methods across both lung nodule segmentation datasets. Fredriksen \textit{et al.} \cite{fredriksen2022teacher} addresses data scarcity utilizing pseudolabels in a similar mixed-UNet architecture. The main difference with our work is the introduction of a 3D bbx instead of a series of 2D bbxs. The proposed model outperforms \cite{fredriksen2022teacher} by 2.98 and 19.44 DSC for the MSD and RAD datasets, respectively. This difference suggests a higher impact on exploiting fine-grained details using high-quality labels over increasing dataset size with lower-quality pseudolabels. The model by \cite{kamal2020lung} is outperformed by a negligible difference of 0.08 DSC. Their implementation makes use of dense connectivity for aggregated feature propagation.

\begin{table}[h]
    \centering
    \begin{tabular}{l|c|c}
        \hline
        \textbf{Experiment} & \textbf{MSD (Dice)} & \textbf{RAD (Dice)}  \\
        \hline
        \makecell[l]{nnUNet 2D\\ full resolution} & $52.68$ \cite{isensee2018nnu} &  $58.50\pm1.57$\\
        \hline
        \makecell[l]{nnUNet 3D\\ full resolution} & $55.84$ \cite{isensee2018nnu} &  $62.68\pm1.03$\\
        \hline
        \makecell[l]{RA-Seg\\(Proposed)} & \textbf{67.25$\pm$3.12} &  \textbf{72.36$\pm$1.69}\\
        \hline
    \end{tabular}
    \caption{nnUNet baseline and proposed ``RA-Seg" DSC results comparison.}
    \label{tab:Final_arch}
\end{table}

As seen in Table \ref{tab:inference}, inference results on the RADGEN test set when training on the MSD dataset slightly outperform the RAD dataset, although the latter is almost six times larger. This result could be attributed to RADGEN tumors' mean volume ($8.22$ $cm^3$) being closer in size to MSD ($21.92$ $cm^3$) compared to RAD tumor size ($75.37$ $cm^3$). Nishio \textit{et al.} \cite{nishio2021lung} generate an artificial dataset using the LUNA16 dataset to finetune a nnUNet model trained on the MSD dataset. The proposed model outperforms by 0.53 DSC when trained on a combined dataset (MSD+RAD) and without pretaining. The increase in performance when using the combined dataset demonstrates the importance of data diversity over data size, as highlighted in \cite{hofmanninger2020automatic}. 

\begin{table}[h]
    \centering
    \begin{tabular}{l|c}
        \hline
        \makecell[l]{\textbf{Experiment} \textbf{(dataset)}} & \textbf{Test (Dice)}\\
        \hline
        \makecell[l]{RA-Seg\\ (MSD)} & $66.73$\\
        \hline
        \makecell[l]{RA-Seg\\ (RAD)} & $66.11$\\
        \hline
        \makecell[l]{RA-Seg\\ (MSD+RAD)} & \textbf{73.54}\\
        \hline
    \end{tabular}
    \caption{DSC results on RADGEN test set. The inference was performed on a model trained on the MSD dataset only, RAD dataset only, and a combined (MSD+RAD) dataset.}
    \label{tab:inference}
\end{table}

\begin{figure}[h]
    \centering
    \includegraphics[scale=0.25]{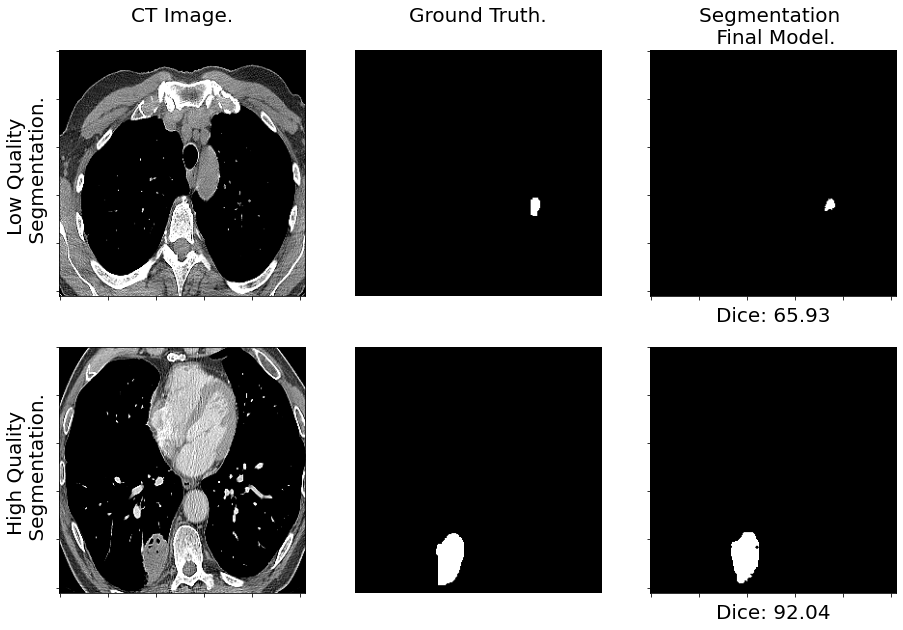}
    \caption{Visual segmentation results of RADGEN dataset finetuned on combined dataset.}
    \label{fig:final_results}
\end{figure}

As a final evaluation, the segmentation method was trained on the RADGEN dataset. The best performance was achieved when utilizing the pretrained weights of the combined dataset, the method achieves a DSC of $75.26\pm4.37$, 1.48 above the 3D nnUNet baseline. These results highlight the extent of transfer learning and further solidify the impact of data diversity over data size. Figure \ref{fig:final_results} shows generated low and high quality segmentations. The low-quality segmentation presents ground glass opacity between 50 and 75\%, according to the available clinical information. DL approaches could benefit from introducing traditional methods to address abnormal nodules as in \cite{mansoor2014generic}.

\subsection{Classification}

Reddy \textit{et al.} \cite{reddy2020analysis} tested the effect of sample size when utilizing dimensionality reduction techniques. In their study, PCA performs better than LDA only when the number of samples increases. Our classification results presented in Table \ref{tab:dim_red} show that LDA outperforms the PCA methodology across all classification algorithms, suggesting the impact of RADGEN data size.

\begin{table}[h]
    \centering
    \begin{tabular}{l|c|c|c}
        \hline
        \textbf{Classifier} & \makecell{\textbf{Baseline}\\(F1-macro)} & \makecell{\textbf{w/PCA}\\(F1-macro)} & \makecell{\textbf{w/LDA}\\(F1-macro)} \\
        \hline
        QDA & 0.447 & 0.446 & 0.936\\
        \hline
        DT & 0.438 & 0.510 & 0.908\\
        \hline
        RF & 0.522 & 0.613 & 0.948\\
        \hline
        SVC & 0.446 & 0.575 & 0.831\\
        \hline
    \end{tabular}
    \caption{F1-macro performance for PCA and LAD dimensionality reduction across multiple classifiers.}
    \label{tab:dim_red}
\end{table}

The ROC-AUC is the most important metric to compare methods since it is threshold and scale-invariant, allowing a fair comparison. As seen in Table \ref{tab:final_class} The best-performing classifier was RF with a ROC-AUC of $0.93\pm0.06$. It provides an improvement over ML approaches of 0.32 \cite{koyasu2020usefulness}, 0.23 \cite{moreno2021radiogenomics}, and 20 \cite{morgado2021machine}. It outperforms SOTA DL approaches by 0.09 \cite{moreno2021radiogenomics} and 0.07 \cite{gui2022air}. Middle-level features from the \textit{Recurrent Block} output were also tested, RF performed best with an AUC-ROC of  $0.91\pm0.03$. Finally, we tested multiple-depth features by concatenating middle and high-level vectors, SVC achieved an AUC-ROC of $0.88\pm0.02$. The proposed method can successfully discriminate deep feature representations for precise prediction of EGFR mutation status.


\begin{table}[h]
    \centering
    \begin{tabular}{l|c|c|c}
        \hline
        \textbf{Classifier} & \textbf{Accuracy} & \textbf{F1-macro} & \textbf{ROC-AUC} \\
        \hline
        QDA & $0.96\pm0.02$ & $0.93\pm0.03$ & $0.90\pm0.05$\\
        \hline
        DT & $0.94\pm0.03$ & $0.91\pm0.05$ & $0.91\pm0.06$\\
        \hline
        RF & \textbf{0.97$\pm$0.02} & \textbf{0.95$\pm$0.04} & \textbf{0.93$\pm$0.06}\\
        \hline
        SVC & $0.90\pm0.11$ & $0.83\pm0.21$ & $0.83\pm0.21$\\
        \hline
    \end{tabular}
    \caption{Accuracy, F1-macro, and ROC-AUC scores across multiple classifiers utilizing LDA dimensionality reduction.}
    \label{tab:final_class}
\end{table}

\section{Conclusion}

This work explores the problem of segmentation of lung cancer and the classification of radiogenomics characteristics. The proposed model reaches SOTA results in lung nodule segmentation thanks to the proposed recurrent and attention modules that help localize the tumor, exploit spatio-temporal features, and finetuning with pretrained weights. The proposed classification methodology for EGFR mutation status outperforms existing approaches without the need for hand-crafted features. The evolution and improvement of radiogenomic techniques have the potential to advance lung cancer precision medicine, preventing over or under-treatment. But community efforts should focus on the development of standardized methods to establish clinical utility.

\section{COMPLIANCE WITH ETHICAL STANDARDS}

This research study was conducted retrospectively using human subject data made available in open access by \cite{aerts2014decoding}, \cite{simpson2019large}, and \cite{bakr2018radiogenomic}. Ethical approval was not required as confirmed by the license attached with the open access data.

\section{Conflicts of Interest}

No funding was received for conducting this study. The authors have no relevant financial or non-financial interests to
disclose

\bibliographystyle{IEEEbib}
\bibliography{refs}

\end{document}